\documentclass[aps,prd,twocolumn,superscriptaddress,nofootinbib]{revtex4}

\usepackage{epsfig}
\usepackage{amssymb}
\usepackage{amsmath}
\usepackage{amsfonts}
\usepackage{hyperref}
\usepackage{color}

\newcommand{\be}{\begin{equation}}
\newcommand{\ee}{\end{equation}}
\newcommand{\bea}{\begin{eqnarray}}
\newcommand{\eea}{\end{eqnarray}}
\newcommand{\bi}{\begin{itemize}}
\newcommand{\ei}{\end{itemize}}
\newcommand{\ben}{\begin{enumerate}}
\newcommand{\een}{\end{enumerate}}

\def\gsim{\mathrel{\rlap{\lower4pt\hbox{\hskip1pt$\sim$}}
    \raise1pt\hbox{$>$}}}         
\def\lsim{\mathrel{\rlap{\lower4pt\hbox{\hskip1pt$\sim$}}
    \raise1pt\hbox{$<$}}}         

\begin{document}

\title{Constraining the Higgs boson width with $ZZ$ production at the LHC}

\author{Fabrizio Caola}
\email{caola@pha.jhu.edu}
\affiliation{Department of Physics and Astronomy, Johns Hopkins 
University, Baltimore, USA}
\author{Kirill Melnikov}
\email{melnikov@pha.jhu.edu}
\affiliation{Department of Physics and Astronomy, Johns Hopkins 
University, Baltimore, USA}

\begin{abstract}
We point out that existing measurements of   $pp \to ZZ$ cross-section at the LHC 
in a broad range of $ZZ$ invariant masses 
allow one to derive a model-independent  upper bound on the Higgs boson width, thanks to strongly enhanced off-shell Higgs contribution. 
Using CMS data 
and considering   events in the interval of $ZZ$ invariant  masses from $100$ to $800~{\rm GeV}$,
we find 
$\Gamma_H \le 38.8 \; \Gamma_{H}^{\rm SM} \approx 163~{\rm MeV}$, at the $95\%$ confidence level.
Restricting   $ZZ$ invariant masses to $M_{ZZ} \ge 300~{\rm GeV}$ range, we estimate that this 
bound can be improved to $\Gamma_H \le  21 \; \Gamma_H^{\rm SM} \approx 88~{\rm MeV}$.
Under the assumption that all couplings of the Higgs boson to Standard Model particles 
scale in a universal way, our  result can be 
translated into an upper limit on the branching fraction of the Higgs 
boson decay to invisible final states. We obtain ${\rm Br}(H \to {\rm inv}) < 0.84~(0.78)$, depending on the range 
of $ZZ$ invariant masses that are used to constrain the width.   We believe that an analysis along these lines should  be 
performed by experimental collaborations in the near future and also in run II of the LHC. We estimate 
that such analyses can, eventually, be sensitive to a Higgs boson width as small as $\Gamma_H \sim 10 \; \Gamma_{H}^{\rm SM}$. 
\end{abstract}

\maketitle

Since the discovery of the Higgs-like  particle  by ATLAS and  CMS   collaborations 
about a year ago \cite{discovery-atlas,discovery-cms}, 
much  has been learned about its properties. 
We know that the mass of the   new particle is around  $126~{\rm GeV}$ \cite{ATLAS-mass,CMS-comb}, 
that its spin-parity  is most likely $0^+$ \cite{ATLAS-spin,cms,CMS-spin}  and that 
its production cross-sections as observed in particular  production and decay channels 
are consistent with Standard Model  expectations  \cite{ATLAS-coupling,CMS-comb}.  
It is customary to translate the latter result into a statement about Higgs boson couplings 
to Standard Model particles but, as it is  well-known, 
such a translation is only possible under the    assumption that the 
Higgs boson width is the same as in the Standard Model (SM). Indeed, since after imposing 
selection cuts   the Higgs boson production  at the LHC can be described 
in a narrow width approximation ~\cite{ Dicus:1987fk, Dixon:2003yb, 
Campbell:2011cu,Kauer:2012hd,Kauer:2013cga}, we can write a production cross-section  for the process $i \to H \to f$ as 
\be
\sigma_{i \to H \to f} \sim  \frac{g_i^2 g_f^2}{\Gamma_H},
\label{eq1}
\ee
where $g_{i,f}$ are the Higgs boson couplings to initial and final states and $\Gamma_H$ is the Higgs boson width.
Therefore, all measured cross-sections can be kept fixed if one simultaneously rescales couplings of the Higgs boson to Standard Model   
particles and the Higgs boson width by appropriate factors.  Indeed, if $g = \xi g_{\rm SM}$ and 
$\Gamma_H = \xi^4 \Gamma_{\rm H,SM}$, the measured Higgs production cross-sections 
in all channels will coincide with expected Standard Model values, $\sigma_{i \to H \to f} = \sigma_{i \to H \to f}^{\rm SM}$.
We conclude that current LHC data allow for   infinitely many solutions for the Higgs couplings to SM particles,  
the Higgs width and the branching fraction of the Higgs boson  to invisible (or so far unobserved)   states.  
To break this degeneracy, independent  measurements of the Higgs boson width  or the Higgs couplings are required. 

Direct  measurement of the Higgs boson width is not possible at a  hadron collider unless $\Gamma_H \gsim {\cal O}(1)~{\rm GeV}$,  
or more than $250$ times larger than its Standard Model value.  The only facility  where   a   direct  measurement of the width 
can be performed  is a future muon collider where by scanning the production cross-section for $\mu^+\mu^- \to H \to X$ 
around $m_H$, the Higgs width can be directly measured to high precision \cite{ Han:2012rb,Conway:2013lca}.  
At any other facility, the Higgs boson width should be obtained indirectly, using 
information on the Higgs  couplings to Standard Model particles or information about   the   Higgs boson branching ratio to invisible final 
states,  provided that such information is available from independent sources.  

A number of ways were suggested to constrain the Higgs couplings and the Higgs branching fraction
into invisible final states. For example, under certain theoretical assumptions 
about electroweak symmetry breaking, one can argue  \cite{gunion} that the SM value of the Higgs boson 
coupling to $W$-bosons provides an {\it upper} bound for all possible $HWW$  couplings.  From this, the upper limit
on the Higgs width $\Gamma_H < 1.43\;   \Gamma_H^{\rm SM} $ is  obtained \cite{Dobrescu:2012td}.
Imposing even stronger constraints on the Higgs couplings to Standard Model particles, one can obtain 
tighter bounds on the Higgs boson width \cite{Barger:2012hv,Djouadi:2013qya}.
Under the assumption of the Standard Model 
production rate for $pp \to ZH$,   the   ATLAS collaboration derives an upper bound on the Higgs branching ratio 
to invisible final state ${\rm Br}(H \to {\rm inv}) < 0.65$ at the   $95\%$ confidence level \cite{atlasinv}. 
A related CMS study with a similar conclusion has also appeared  recently \cite{cmsinv}. 

On the other 
hand, it is more difficult to obtain model-independent constraints on the Higgs boson couplings. 
It was suggested in  Ref.~\cite{Dixon:2013haa} to use differences  in the measured values of the Higgs boson masses 
in $\gamma \gamma$ and $ZZ$ channels, caused by the interference of $gg \to H \to \gamma \gamma $ and 
$gg \to \gamma \gamma $ amplitudes, as a tool to constrain the product 
of $Hgg$ and $H\gamma \gamma$ couplings,  independent of the Higgs boson width.  Once the couplings are measured, 
one can derive the value of the Higgs boson width from the narrow 
width cross-section, see Eq.(\ref{eq1}). 

The purpose of this paper  is to point out that a  constraint on the product of 
$Hgg$ and $HZZ$ couplings  and the resulting  model-independent constraint on the 
Higgs boson width can be obtained from the observed number of $ZZ$ events at the LHC {\it above } the Higgs boson mass peak 
in the $pp \to ZZ$ process. Interestingly, this can already be done with the current data. The main
reason for that is an enhanced contribution to the Higgs signal from invariant masses above the $ZZ$ threshold, 
as was first pointed out in Ref.~\cite{Kauer:2012hd}. Interestingly, useful limits on the Higgs width
can already be derived with the current data. 
To show how this works, we  recall how Eq.(\ref{eq1}) is obtained.  We  focus on the
$H \to ZZ \to ee \mu \mu$ final state and write the production cross-section as a function of the invariant mass
of four leptons $M_{4l}$   
\be
\frac{{\rm d} \sigma_{pp \to H \to ZZ}}{{\rm d} M_{4l}^2} 
\sim \frac{g_{Hgg}^2 g_{HZZ}^2}{(M_{4l}^2 - m_H^2)^2 + m_H^2 \Gamma_H^2}.
\label{eq2} 
\ee

The  total cross-section receives the dominant contribution from 
the resonant region $M_{4l}^2 - m_H^2 \sim m_H \Gamma_H$,
where integral of Eq.(\ref{eq2}) gives  Eq.(\ref{eq1}). However, the total cross-section also receives off-peak contributions
from  larger or smaller invariant masses, where Eq.(\ref{eq2}) 
is still proportional 
to squares of $Hgg$ and $HZZ$ couplings
{\it but it  is independent  of} $\Gamma_H$.
  
Suppose now that in Eq.(\ref{eq2}), the product of coupling constants 
$c_{gZ} = g_{Hgg}^2 g_{HZZ}^2$ and the width $\Gamma_H$
are scaled by a common  factor  $\xi$ and that this factor  is still sufficiently small  to make 
the narrow width approximation applicable.
Under this circumstance, the resonance contribution remains unchanged and is  given 
by Eq.(\ref{eq1}), while the off-shell  contribution from 
the region $M_{4l}^2 \gg m_H^2$   increases {\it linearly} with $\xi$ and can, therefore, be bounded from 
above by the total number of events observed in  $pp \to ZZ$ process above the Higgs boson peak in the $ZZ$ invariant 
mass spectrum. This is the main idea behind this paper. 

\begin{figure}
\includegraphics[scale=0.3]{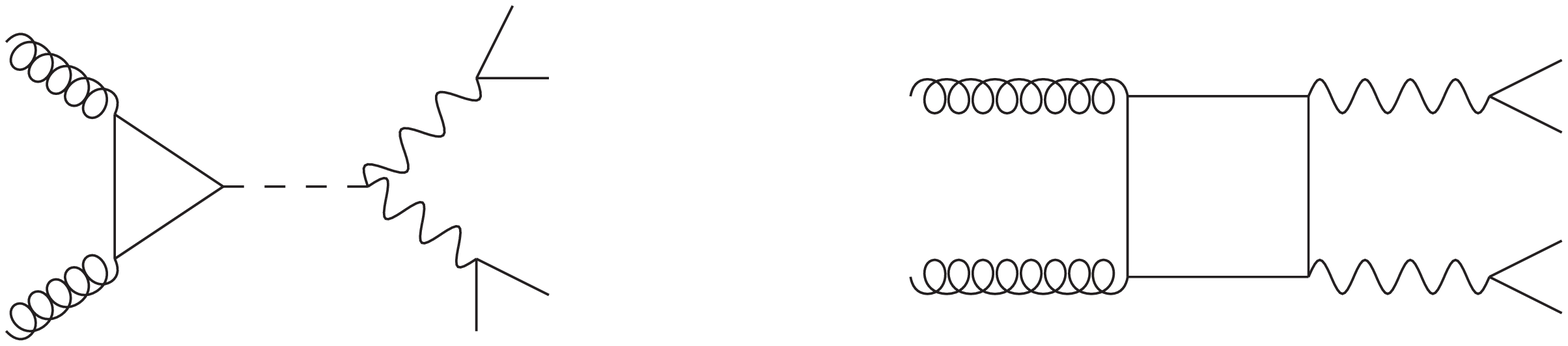}
\caption{Sample signal (left) and background $gg \to ZZ$ (right) diagrams 
for the process $pp\to ZZ \to 4l$. The two amplitudes can interfere.}\label{fig1}
\end{figure}

There are two sources of  Higgs-related $ZZ$ events off the peak. One is the off-shell production 
of the Higgs boson followed by its decay to $ZZ$ final states. The second source of events   is   the  
interference between $gg \to H \to ZZ$  and $gg \to ZZ$ 
amplitudes, see Fig.~\ref{fig1}. The interference exists, but is  numerically irrelevant {\it in the peak} ~\cite{Kauer:2012hd,Kauer:2013cga}  
while,  as we show below,
it significantly changes the number of expected Higgs-related events off the peak. 
We account for both of these effects in the following discussion. To estimate  the number of Higgs events 
in $gg \to H \to ZZ$, including the interference,  we use 
the program {\sf gg2VV} described in Refs.~\cite{Kauer:2012hd,Binoth:2008pr}. 

To calculate the number of Higgs-related events that are expected off peak, we compute  $7$ and $8~{\rm TeV}$ 
production cross-sections 
for $pp \to H \to ZZ \to e^+e^-\mu^+\mu^-$ at {\it leading order} in perturbative QCD  
requiring that the invariant 
mass of four leptons is either smaller or larger than $130~{\rm GeV}$.  We refer to the former case as the 
``on peak'' cross-section and to the latter case as   the ``off peak'' one .

We employ the 
CMS selection cuts \cite{cms}  requiring  $p_{\perp,\mu} > 5~{\rm GeV}$, $p_{\perp, e} > 7~{\rm GeV}$, 
$|\eta_{\mu}| < 2.4$, $|\eta_e|< 2.5$, $M_{l_- l_+} > 4~{\rm GeV}$, $M_{4l} > 100~{\rm GeV}$. In addition, the 
transverse momentum of the hardest (next-to-hardest) lepton should be larger than $20\;(10)~{\rm GeV}$,
the invariant mass of a pair of same-flavor leptons closest to the $Z$-mass should 
be in the interval $40 < m_{ll} < 120~{\rm GeV}$ and the invariant mass of the other pair should be 
in the interval $12-120~{\rm GeV}$.  We also take 
the Higgs boson mass to be $126~{\rm GeV}$, and set renormalization and factorization scales to $m_H/2$.   

\begin{table}[t]
\begin{center}
\begin{tabular}{|c|c|c|c|c}
\hline
~~~~{\rm Energy} ~~~& ~~~~~~~$\sigma_{\rm peak}^{\rm H}$ ~~~~~~~& $ ~~~~~~~\sigma_{\rm off}^{H}$ ~~~~~~~~~~& ~~~~$\sigma_{\rm off}^{\rm int}~~~~~~~$ 
 \\
\hline\hline
{\rm 7 TeV} &  0.203  &  0.044  &   -0.108   \\
\hline
{\rm 8 TeV} & 0.255  &   0.061   &  -0.166   \\
\hline
$N_{2e2\mu}^{\rm SM} $ & 9.8  & 1.73  &  -4.6  \\
\hline \hline
$N_{\rm tot}^{\rm SM} $ & 21.1  &   3.72   &  -9.91  \\
\hline\hline
\end{tabular}
\vspace*{0.5cm}
\caption{Fiducial cross-sections for $pp \to H \to ZZ \to 2e2\mu$ in fb, and the corresponding number of events expected 
for integrated luminosities  $L_7 = 5.1~{\rm fb}^{-1}$ at $7~{\rm TeV}$ and 
$L_8 = 19.6~{\rm fb}^{-1}$ at $8~{\rm TeV}$. All cross-sections are computed with leading order 
MSTW 2008 parton distribution functions \cite{Martin:2009iq}. The renormalization and factorization 
scales are set to $\mu = m_H/2$.  The peak cross-section is defined with the cut $M_{4l} < 130~{\rm GeV}$,
while off-peak and interference cross-sections are defined with the cut $M_{4l} > 130~{\rm GeV}$. 
The total number of events in the last row includes contributions from $4e$ and $4\mu$   channels. 
The number of events is obtained using procedures outlined in the text. 
}
\label{tab1}
\end{center}
\end{table}

The corresponding cross-sections for the Higgs signal on and off
the peak as well as the interference contributions to cross-sections are shown in Table~\ref{tab1}. The number 
of $2e2\mu$  events in that Table is computed  starting from the number of on-peak events  reported in 
Table~I of Ref.~\cite{cms}. According to Table~I in \cite{cms},
the CMS collaboration expects $9.8$ Higgs-related 
events in the $ee\mu \mu$ channel {\it on the peak}.\footnote{This number of events is a combination 
of $gg \to H$ ($88\%$), weak boson fusion ($7\%$) and 
$VH$ production ($5\%$). Although a detailed study of the channels besides $gg \to H $ is beyond the scope of this paper, 
we believe that they  will contribute to the number of high-mass $ZZ$ events 
in a way that is similar to $gg \to H \to ZZ$; for this reason we decided to keep the number of events in the peak unchanged when 
performing numerical   estimates.} We estimate the number of Higgs-related events for $M_{4l} > 130~{\rm GeV}$ 
by taking ratios of cross-sections weighted with luminosity factors. 
We also include  additional suppression factor due to the fact that the appropriate scale choice for the strong coupling 
constant in $gg \to H^* \to ZZ$ is the invariant mass of the $Z$ boson pair divided by two,  rather than $m_H/2$,  
as appropriate for the on-shell cross-section \cite{Anastasiou:2002yz}.
We take $300~{\rm GeV}$ as a typical value of the invariant mass for Higgs-related events produced off the peak. The corresponding 
suppression factor is then given by $\eta = (\alpha_s(150~{\rm GeV})/\alpha_s(m_H/2))^2 \approx 0.75$. We find
\be
 N_{2e2\mu}^{H,\rm off} = 9.8 \times \eta  
\frac{ L_7 \sigma_{\rm off}^{H}(7)  + L_8 \sigma_{\rm off}^{H}(8) }{ L_7 \sigma_{\rm peak}^{H}(7) 
+ L_8 \sigma_{\rm peak}^{H}(8) }
  \approx 1.73,
\ee
where we use  the integrated luminosities $L_7 = 5.1~{\rm fb}^{-1}$ at $7~{\rm TeV}$ and 
$L_8 = 19.6~{\rm fb}^{-1}$ at $8~{\rm TeV}$.

We combine this estimate with results  for other  lepton channels by similarly rescaling CMS data on $4e$ and $4\mu$, and conclude that 
$3.72$ four-lepton events produced by decays of an off-shell Higgs boson 
can be expected in the  current data.   Repeating this calculation  with the interference   contribution, 
we find that 
$-9.91$ events are expected.  Since  cross-sections that we use are computed in the leading order QCD 
approximation and 
do not include any detector effects, one may 
wonder if the number of events estimated using them is reliable.
  While a detailed answer to this question requires 
careful studies, we believe that, by taking ratios of cross-sections, accounting for  the dominant effects of the running of the 
strong coupling constant when relating  on- and off-peak events 
and by normalizing our computation  to the CMS number  of the expected Higgs  events in the peak, 
we obtain estimates for the off-peak number of events that are sufficiently reliable  
for the purposes of this paper.\footnote{We note that by rescaling both  off-peak and interference contributions  in the
same way, we implicitly assume that QCD corrections  to  the signal  and the interference are comparable. This
is supported by the analysis of higher-order corrections to the interference in $pp \to H \to W^+W^-$ process
reported  in~\cite{Bonvini:2013jha}.}

We note that the estimated number of events in  Table~\ref{tab1} looks quite striking for two reasons.
 The first one is that the off-shell contributions related 
to $gg \to H \to ZZ$ are {\it large}; the off-peak cross-section is close  to twenty percent   of  
the peak cross-section. 
This large off-peak contribution in $ZZ$ final state was 
first emphasized in  Ref.~\cite{Kauer:2012hd}. 
It was explained as the consequence of a relatively large 
probability to produce the Higgs boson with the off-shellness 
larger than  $2 m_Z$ where decays to longitudinally-polarized $Z$-bosons rapidly become 
important and compensate for the decrease in the cross-section caused  by the off-shell Higgs propagator. 
This leads to a contribution to the invariant mass distribution Eq.(\ref{eq2}) which, although small, extends
over a large invariant mass range $2 m_Z \lsim M_{4l} \lsim 800~{\rm GeV}$ and gives  rise to a sizable contribution
to the total cross-section. 
The second reason is due to a large destructive interference. Note, however, that the interference is an {\it off-peak} 
phenomenon; it does not contribute to the peak cross-section to a very good approximation~\cite{Kauer:2012hd,Kauer:2013cga}. 

The expected number of Higgs-related events shown in Table~\ref{tab1}
refers to the Standard Model. 
Relaxing this assumption by allowing for correlated changes in the Higgs couplings and the Higgs boson 
width,  so that the number of events in the peak remains intact, we write the number of off-peak events as 
\be
 N_{4l}^{\rm off} =  3.72 \times \frac{\Gamma_H}{\Gamma_H^{\rm SM}} - 9.91 \times \sqrt{\frac{\Gamma_H}{\Gamma_H^{\rm SM}}}.
\label{eq5}
\ee

For $\Gamma_H \gg  \Gamma_H^{\rm SM}$, 
we can interpret Eq.(\ref{eq5}) as an additional source of $ZZ$ events in the current data; these $ZZ$ events 
are broadly distributed over   a   large invariant mass range, roughly from 
the  $ZZ$ threshold up to the highest 
$ZZ$ invariant masses of order 800~{\rm GeV}. Therefore, as the first step, we can
look at the total number of $ZZ$-events in the current data and ask how many additional events can be tolerated 
given the number of observed events and the current uncertainty on the number of expected events. 
CMS currently observes $451$ events in the $pp \to ZZ \to 4l $ channel, while $432 \pm 31$ events are expected \cite{cms}. The 
expected number of events does not include the off-shell Higgs production and the off-shell interference. Therefore, 
we estimate the total number of events that are  expected if the Higgs couplings and width
differ from the Standard Model using the following equation
\be
N_{\rm exp} = 432 +  3.72 \times \frac{\Gamma_H}{\Gamma_H^{\rm SM}} - 9.91 \times \sqrt{\frac{\Gamma_H}{\Gamma_H^{\rm SM}}}
\pm 31 ,
\ee
where we assume that the sign of the interference is the same as in the Standard Model.
Note that we obtain the above error estimate by adding 
errors  for the $4e$, $4\mu$ and $2e 2\mu$ channels reported in Ref.~\cite{cms} 
 in quadratures, assuming that they are uncorrelated. While not exact, 
this is also not an unreasonable assumption,\footnote{Note that errors for the expected number of 
background events for all channels in Table~I of Ref.~\cite{cms}  
are of the same order as the square root of the expected number of events reported there.}
but a detailed analysis of error correlations is beyond 
the scope of this paper.  

Requiring that the expected and observed numbers of events are within two standard deviations  from each other, we derive an upper limit 
on $\Gamma_H$ at the $95\%$ confidence level. We find 
\be
\Gamma_H \le 38.8 \; \Gamma_H^{\rm SM}  \approx 163~{\rm MeV},
\ee
where we used $\Gamma_H^{\rm SM}  \approx 4.2~{\rm MeV}$  \cite{Djouadi:1997yw}.\footnote{We note that, if we add the errors for the number 
of expected events in the $4e$, $4\mu$ and $2e 2\mu$ channels {\it linearly}, the $95\%$ confidence level limit 
for the width will degrade to $\Gamma_H \le 52 \; \Gamma_H^{\rm SM}$. }

\begin{table}[t]
\begin{center}
\begin{tabular}{|c|c|c|c|}
\hline
~~~~{\rm Energy}~~~~~ & ~~~~~~$\sigma_{\rm peak}^{\rm H}$ ~~~~~& ~~~~~~$ \sigma_{\rm off}^{H}$~~~~~~ &~~~~~~ $\sigma_{\rm off}^{\rm int}$~~~~~ \\
\hline\hline
{\rm 7 TeV} &  0.203  &    0.036   &   -0.046  \\
\hline
{\rm 8 TeV} & 0.255  &   0.049    & -0.10 \\
\hline
$N_{2e2\mu}^{\rm SM} $ & 9.8  & 1.39  &  -2.71 \\
\hline \hline
$N^{\rm SM}_{\rm tot} $ & 21.1  &   2.99    &    -5.84  \\
\hline\hline
\end{tabular}
\vspace*{0.5cm}
\caption{Same of Table~\ref{tab1}, but with the cut $M_{4l} > 300$~GeV applied to the off-peak
cross-section and interference. See text for details.
}
\label{tab2}
\end{center}
\end{table}

The upper limit on the Higgs boson width can be turned into an upper limit on the branching fraction for the Higgs 
boson decay into invisible final states.  To this end, we
write  
\be
\Gamma_H = \Gamma_{\rm inv} + \sum \limits_{i \in {\rm vis}} \Gamma_i,
\ee
where the sum extends over all visible channels. We note that $\Gamma_{i \in {\rm vis}} \sim g_i^2$, 
and that ratios  $g_i^2 g_f^2/\Gamma_H$ should be equal to their  Standard Model values, 
to keep all narrow-width Higgs boson production cross-sections to be the same as in the Standard Model.
Assuming that all Higgs couplings to SM particles differ by identical factors relative to their Standard Model 
values,  we find that the Higgs boson width and the branching fraction to invisible final states satisfy
the following constraint
\be
\Gamma_H \left ( 1 - {\rm Br}_{\rm inv} \right )^2 = \Gamma_H^{\rm SM}.
\label{eq15}
\ee
This constraint translates into an  upper limit on ${\rm Br}_{\rm inv}$ 
\be
{\rm Br}_{\rm inv} = 1 - \sqrt{\Gamma_H/\Gamma_H^{\rm SM}} < 0.84.
\ee

Can the above analysis be improved? We believe that there is, most likely, an affirmative answer to this question. 
To show this, we note that an upper bound on the Higgs 
width was derived by using the total number of $pp \to ZZ$ events observed in a broad range of four-lepton invariant masses. 
However, this may not be an  optimal mass range  since  the invariant mass   distribution   of the four-lepton events produced 
in the ``decays'' of the off-shell Higgs   boson   is almost flat.  To illustrate this point, 
we repeat the  above  analysis but now select events where  the invariant mass of four leptons is larger than 
$300~{\rm GeV}$.  The corresponding leading order cross-sections are shown in  Table~\ref{tab2}. By comparing 
Tables~\ref{tab1} and \ref{tab2}, it is clear that the off-shell production decreases by a smaller amount than the interference. 
The observed number of events for $M_{4l} > 300~{\rm GeV}$ is $N_{\rm obs} = 87$ and the expected number of events 
is estimated to be $N_{\rm exp} = 70.7$ without off-shell Higgs production and the interference \cite{cms}.  It is not possible 
for us 
to obtain the error estimate for  expected number of events from the CMS paper  \cite{cms}; we therefore 
take $\delta N_{\rm exp} = 10$ which is about 15 percent of $N_{\rm exp}$.  Repeating the same analysis 
as in the case of the full mass range, we find an improved  $95\%$ confidence level limit 
on the Higgs boson width 
\be
\Gamma_{H} \le 21~\Gamma_H^{\rm SM} \approx 88~{\rm MeV}. 
\ee

Further refinements  should, therefore, include a careful selection of the   invariant mass window and, perhaps, the use 
of angular correlations of four lepton momenta 
to disentangle $gg \to H \to ZZ$ off-peak events from $q \bar q \to ZZ$ background. Such 
angular correlations are already used by the CMS collaboration \cite{cms} to improve their measurement in the Higgs 
peak  region; it is 
probably straightforward to apply these techniques  off the peak as well. We note that  polarization effects 
may play a more substantial role 
at high-invariant masses since $Z$ bosons that are  produced in decays of the off-shell Higgs boson 
are, most likely, longitudinally polarized.

With increased luminosity, one can expect the error on the number of $ZZ$ events to be  dominated by systematic uncertainties;
we will optimistically assume that this uncertainty will, eventually, become as  small as  $3\%$. This may require extending 
existing theoretical computations for $pp \to ZZ$ to NNLO QCD but this appears  to be a realistic 
target on a few years time-scale; see e.g. Ref.~\cite{Gehrmann:2013cxs} as an example of recent 
progress.   If such an error is reached and about half of the background events 
are rejected,  the $95\%$ confidence level upper limit on the Higgs boson width $\Gamma_H \lsim 5-10 ~ \Gamma_H^{\rm SM} = 20-40~{\rm MeV}$ 
may, eventually, be obtained.  This appears to be the ultimate limit of what can be reached with the methods 
that are advocated in this paper.  

In conclusion, we suggested that the total Higgs boson width 
can be constrained  in a model-independent way by studying the $ZZ$ events 
off the Higgs boson invariant-mass peak.  We pointed out that already with the current data one can put a  $95\%$ confidence 
limit $\Gamma_H \le 20 - 38 ~\Gamma_{H}^{\rm SM}$ depending on the four-lepton invariant mass range chosen for 
the analysis. We also note that if the interference contribution in Eq.(\ref{eq5}) changes sign and becomes constructive,  bounds on the 
Higgs width become much stronger, $\Gamma_H \le 7-13 \; \Gamma_H^{\rm SM}$.
While we believe that our estimates are sufficiently accurate, 
the present  study  is crude and ignores the many details of experimental event selection. We tried to mitigate that 
by normalizing our calculations to the number of Higgs boson events that CMS collaboration 
expects  to observe in the peak. However, it will be 
best if experimental collaborations perform a detailed analysis of $ZZ$ events at  
high invariant masses and, as suggested in this paper,  derive model-independent 
constraints on the Higgs boson width.

{\bf Acknowledgments} 
We are grateful to A.~Gritsan,  F. Krauss  and A. Whitbeck for useful discussions and encouragement. 
We thank A.~Gritsan for many detailed comments on the manuscript. 
We are grateful  to N.~Kauer for discussions on the importance of off-shell effects, and for help with the program {\sf gg2VV}.  
This research is partially supported by US NSF under Grant No. PHY-1214000.

\end{document}